\documentclass[a4paper]{jpconf}
\usepackage{graphicx}
\begin{document}
\title{On five-dimensional massive (bi-)gravity}

\author{Tuan Q Do}

\address{Faculty of Physics, VNU University of Science, Vietnam National University, Hanoi 120000, Vietnam}

\ead{tuanqdo@vnu.edu.vn}
%%%%%%%%%%%%%%%%%%%%%%%%%%%%
\begin{abstract}
Main results of our recent investigations on five-dimensional scenarios of massive (bi-)gravity will be summarized in this article. In particular, we will show how to construct higher dimensional massive graviton terms from the characteristic equation of square matrix, which is a consequence of the Cayley-Hamilton theorem. Then, we will show whether massive graviton terms of five-dimensional massive (bi-)gravity behave as effective cosmological constants for a number of physical metrics compatible with fiducial ones such as the Friedmann-Lemaitre-Robertson-Walker, Bianchi type I, and Schwarzschild-Tangherlini metrics. Finally, we will show the corresponding cosmological solutions for the five-dimensional massive (bi-)gravity.
\end{abstract}

\section{Introduction}\label{sec1}
Massive gravity, in which a graviton is assumed to have tiny but non-vanishing mass, has a rich and long story since a seminal paper by Fierz and Pauli (FP)~\cite{FP}. In particular, it was shown by van Dam and Veltman \cite{vanDam:1970vg} and Zakharov~\cite{Zakharov:1970cc} that the FP theory cannot recover the general relativity by Einstein in the massless limit. However, Vainshtein showed in \cite{Vainshtein:1972sx} that non-linear extensions of FP theory might resolve the so-called van Dam-Veltman-Zakharov (vDVZ) discontinuity problem. Unfortunately, Boulware and Deser (BD) claimed, shortly after the paper of Vainshtein, in \cite{BD} that a unexpected ghost might arise from the sixth mode of graviton in non-linear FP theory. Therefore, building a non-linear but ghost-free massive gravity has been a great challenge to physicists for  quite a long time. Indeed, many attempts to build up a ghost-free non-linear massive gravity have been proposed, e.g., in \cite{NAH}. However, most of them have not turned out to be successful approaches except a non-linear massive gravity constructed by de Rham, Gabadadze, and Tolley (dRGT)~\cite{RGT} as a generalization of FP massive gravity. As a result, the most important property of the dRGT theory is that it has been proved  to be free of the BD ghost by different approaches  whatever the form of reference (or  fiducial) metric \cite{proof}.

Consequently, many cosmological and physical aspects  of the dRGT theory have been investigated extensively. For example,  the dRGT theory has been expected to provide us an alternative solution to the cosmological constant problem associated with the expanding of our universe. Black holes such as the Schwarzschild, Kerr, and charged black holes along with anisotropic metrics such as the Bianchi type I metric have also been investigated in the context of dRGT theory. More interestingly, the massive gravity and condensed matter theories have met each other in  the so-called holographic massive gravity. For more detailed information, one can read a recent interesting review paper by de Rham \cite{review}. 

Along with the above investigations, many extensions of the non-linear massive gravity have also been carried out~\cite{review}. One of them is the massive bi-gravity (or bi-metric gravity) proposed by Hassan and Rosen~\cite{SFH}, which seems to be one of the most interesting generalizations of dRGT theory due to the change that the reference metric is introduced to be fully dynamical as the physical one. Note that the reference metric in the dRGT theory has been fixed as a non-dynamical metric. Along with the original (nonlinear) massive gravity, the massive bi-gravity has also been discussed extensively, see, for example, a recent interesting review paper~\cite{review-bigravity} for more details. 

Theoretically, inspired by the well-known higher dimensional theories, e.g., the string and super-gravity theories, one could think of higher dimensional scenarios of both massive gravity and bi-gravity. However, we have observed that many published papers on massive (bi-)gravity have focused only on four-dimensional spacetimes, in which only the first three graviton terms ${\cal L}_2$, ${\cal L}_3$, and ${\cal L}_4$ exist. Besides, there are a few papers discussing higher dimensional massive gravity and bi-gravity~\cite{higher-bigravity,higher-bigravity-more}. However, they have not studied particular metrics such as the well-known Friedmann-Lemaitre-Robertson-Walker~\cite{5d-FRW}, Bianchi type I~\cite{5d-bianchi}, and  Schwarzschild-Tangherlini~\cite{5d-sch} metrics for higher dimensional scenarios of massive (bi-)gravity. Hence, we would like to investigate cosmological implications of both five-dimensional massive gravity and bi-gravity  in \cite{TQD,TQD1}. In particular, we are interested in examining whether the metrics mentioned above act as cosmological solutions of  both five-dimensional massive gravity and bi-gravity. To do this, we first construct the graviton term ${\cal L}_5$, which vanishes in all four-dimensional spacetimes but does survive in any five- (or higher) dimensional one, by using the consequence of the Cayley-Hamilton theorem in algebra~\cite{CH-theorem}. Then, we derive and solve constraints equations associated with the existence of reference metric to calculate the value(s) of massive graviton terms ${\cal L}_i$. Finally, we solve the corresponding field equations of physical metric to get the desired cosmological solutions.

The present article will be regarded as a summary of investigations on both five-dimensional massive gravity and bi-gravity. It will be organized as follows. A short introduction of our study has been written in the section~\ref{sec1}. Basic setup and simple cosmological solutions of five-dimensional massive (bi-)gravity will be shown in the sections~\ref{sec2} and \ref{sec3}, respectively. Finally, conclusions  will be given in the section~\ref{sec4}. 

%%%%%%%%%%%%%%%%%%%%%%%%%%%%%
\section{Basic setup of five-dimensional massive (bi-)gravity} \label{sec2}
\subsection{Cayley-Hamilton theorem and massive graviton terms}
First, we would like to mention an action of the dRGT theory in four-dimensional spacetimes~\cite{RGT}:
\begin{equation} \label{action}
S_{4d} = \frac{M_p^2}{2}\int {d^4 x \sqrt {-g}} \Bigl\{ {R +m_g^2\left({{\cal L}_2+\alpha_3 {\cal L}_3+\alpha_4{\cal L}_4}\right)} \Bigr\},
\end{equation}
where $m_g \neq 0$ the graviton mass, $M_p$  the Planck mass,  $\alpha_{3,4}$ free parameters, and ${\cal L}_i$ ($i=2-4$)  the massive graviton terms (or 
interaction terms) given by
\begin{eqnarray}
 \label{eq1}
&&{\cal L}_2 = [{\cal K}]^2- [{\cal K}^2], \\
 \label{eq2}
&&{\cal L}_3 =   \frac{1}{3}  [{\cal K}]^3-[{\cal K}] [{\cal K}^2]+\frac{2}{3}[{\cal K}^3] , \\
 \label{eq3}
&&{\cal L}_4 =   \frac{1}{12}[{\cal K}]^4-\frac{1}{2}[{\cal K}]^2 [{\cal K}^2]+\frac{1}{4} [{\cal K}^2]^2+\frac{2}{3}[{\cal K}][{\cal K}^3]-\frac{1}{2}[{\cal K}^4].
\end{eqnarray}
Here, square brackets used in the definition of  ${\cal L}_i$ are understood as the trace of matrix~\cite{RGT}, e.g.,
\begin{equation} \label{eq4}
 [{\cal K}] \equiv tr{\cal K}^\mu{ }_{\nu};~ [{\cal K}]^2 \equiv \left({{tr}{\cal K}^\mu{ }_\nu}\right)^2;  ~[{\cal K}^2] \equiv {tr}{\cal K}^\mu{ }_\alpha {\cal K}^\alpha{ }_\nu,
\end{equation}
with the definition of $ {\cal K}^\mu{ }_\nu$ is given by
\begin{equation} 
 {\cal K}^\mu{ }_\nu \equiv  \delta^\mu{ }_\nu - \sqrt{g^{\mu\alpha} f_{ab}\partial_\alpha \phi^a \partial_\nu \phi^b },
\end{equation}
where $g_{\mu\nu}$ the physical metric, $f_{ab}$ the reference (or  fiducial) metric, and $\phi^a$ ($a=0-3$) the St\"uckelberg scalar fields added to give a manifestly diffeomorphism invariant description~\cite{NAH,RGT}. It is noted that unlike the physical metric the fiducial metric in the dRGT massive gravity is non-dynamical. Therefore, its field equations will be algebraic rather than differential. For detailed investigations on physical and cosmological aspects of the dRGT theory, see an interesting review paper by de Rham~\cite{review}. 

However, if we introduce  the fiducial metric to be fully dynamical as the physical metric, we will have the corresponding extension of the dRGT theory called a massive bi-gravity (or bi-metric gravity), whose action is defined as follows~\cite{SFH}  
\begin{equation} \label{action1}
S _{4d,bi}= M_g^2 \int {d^4 } x\sqrt { g} R(g)+ M_f^2 \int {d^4 } x\sqrt {f} R(f) +2m^2 M_{{eff}}^2 \int {d^4 } x\sqrt { g} \Bigl( {\cal U}_2 +\alpha_3 {\cal U}_3 +\alpha_4 {\cal U}_4 \Bigr),
\end{equation}
where ${\cal U}_i = {\cal L}_i/2$, while $R(g)$  and $R(f)$ denote the corresponding Ricci scalar of the physical metric $g_{\mu\nu}$ and reference metric $f_{\mu\nu}$, respectively. Additionally,  $M_{{eff}}$ is an effective Planck mass defined as follows \cite{SFH}
\begin{equation}
M_{{eff}}^2 = \left(\frac{1}{M_g^2}+ \frac{1}{M_f^2}\right)^{-1},
\end{equation}  
with $M_g$ and $M_f$ are two other Planck masses of $g_{\mu\nu}$ and $f_{\mu\nu}$, respectively. It is also noted that  ${\cal K}^\mu{ }_\nu \equiv {\delta}^\mu{ }_\nu- \sqrt{g^{\mu\sigma}f_{\sigma\nu}}$ in the context of bi-gravity theory. Of course, due to the dynamical property (or the existence of Ricci scalar $R(f)$), field equations of reference metric in bi-gravity will no longer be algebraic but differential as that of physical metric. For an interesting review of recent developments of bi-gravity theory, see~\cite{review-bigravity}.

As a result, we have been able to show  that the four-dimensional graviton terms, ${\cal L}_2$, ${\cal L}_3$, and ${\cal L}_4$,  can be reconstructed by applying the well-known Cayley-Hamilton theorem in linear algebra~\cite{CH-theorem} for a $4\times 4$ matrix ${\cal K^\mu{ }_\nu}$. Mathematically, this theorem states that any $n\times n$ matrix $K$ must obey its characteristic equation:
\begin{eqnarray} 
{\cal P}(K) \equiv K^n- {\cal D}_{n-1}K^{n-1}+{\cal D}_{n-2}K^{n-2}-\ldots + (-1)^{n-1} {\cal D}_1  K+ (-1)^{n} \det(K) I_n =0,
\end{eqnarray}
where ${\cal D}_{n-1} = tr K \equiv [K]$ and ${\cal D}_{n-j}$ ($2\leq j\leq n-1$) are coefficients of the characteristic polynomial~\cite{CH-theorem}. It is straightforward to show that the forms of the first three graviton terms, ${\cal L}_2$, ${\cal L}_3$, and ${\cal L}_4$ look similar to the determinants of $K_{2\times 2}$, $K_{3\times 3}$, and $K_{4\times 4}$, respectively. This is indeed a key observation to construct any higher dimensional graviton terms, ${\cal L}_{i>4}$~\cite{review-bigravity,higher-bigravity,higher-bigravity-more,TQD}. As a result, it turns out that ${\cal L}_i = 2\det {\cal K}_{i \times i}$ with $i=2-n$.  For example, we have been able to build up five-, six-, and seven-dimensional graviton terms as specific demonstrations~\cite{TQD,TQD1}:
\begin{eqnarray}\label{L5}
\frac{{\cal L}_5}{2} &=&\frac{1}{120}\Bigl\{ [{\cal K}]^5 -10 [{\cal K}]^3 [{\cal K}^2] +20[{\cal K}]^2 [{\cal K}^3]   -20  [{\cal K}^2][{\cal K}^3] +15[{\cal K}]  [{\cal K}^2]^2 -30[{\cal K}] [{\cal K}^4]  +24 [{\cal K}^5] \Bigr\}, \nonumber\\
\end{eqnarray}
\begin{eqnarray}\label{l6}
\frac{{\cal L}_6}{2} &=& \frac{1}{720} \Bigl\{ [{\cal K}]^6 -15[{\cal K}]^4 [{\cal K}^2]+40[{\cal K}]^3 [{\cal K}^3] - 90 [{\cal K}]^2 [{\cal K}^4]  +45 [{\cal K}]^2 [{\cal K}^2]^2  -15 [{\cal K}^2]^3 +40 [{\cal K}^3]^2    \nonumber\\
&&  -120  [{\cal K}^3] [{\cal K}^2] [{\cal K}]+90[{\cal K}^4] [{\cal K}^2]   +144 [{\cal K}^5] [{\cal K}]-120 [{\cal K}^6] \Bigr\},
\end{eqnarray}
\begin{eqnarray}
\label{L7}
\frac{{\cal L}_7}{2} &=& \frac{1}{5040} \Bigl\{[{\cal K}]^7 -21 [{\cal K}]^5[{\cal K}^2]+70 [{\cal K}]^4[{\cal K}^3] -210[{\cal K}]^3 [{\cal K}^4] +105[{\cal K}]^3[{\cal K}^2]^2 -420[{\cal K}]^2[{\cal K}^2][{\cal K}^3]   \nonumber\\
&&+504[{\cal K}]^2[{\cal K}^5] -105[{\cal K}^2]^3[{\cal K}]+210[{\cal K}^2]^2 [{\cal K}^3]-504[{\cal K}^2][{\cal K}^5] +280[{\cal K}^3]^2[{\cal K}] -420[{\cal K}^3] [{\cal K}^4] \nonumber\\
&&+630[{\cal K}^4] [{\cal K}^2] [{\cal K}]  -840[{\cal K}^6][{\cal K}]+720[{\cal K}^7] \Bigr\}.
\end{eqnarray}
 %%%%%%%%%%%%%
\subsection{Five-dimensional massive gravity}
For simplicity, we would like to investigate a five-dimensional massive gravity with the following action~\cite{TQD}:
\begin{equation} 
S_{5d} = \frac{M_p^2}{2}\int {d^5 x \sqrt {-g}} \Bigl\{ {R +m_g^2\left({{\cal L}_2+\alpha_3 {\cal L}_3 +\alpha_4 {\cal L}_4 +\alpha_5 {\cal L}_5} \right)} \Bigr\},
\end{equation}
here  $\alpha_{5}$ is an additional field parameter associated with the massive graviton term ${\cal L}_5$ defined in Eq. (\ref{L5}). As a result, the corresponding Einstein field equations for the physical metric turn out to be~\cite{higher-bigravity,higher-bigravity-more,TQD}
\begin{equation} \label{Einstein-5d}
\left({R_{\mu\nu}-\frac{1}{2}Rg_{\mu\nu}}\right)+m_g^2 \left({ { X}_{\mu\nu}+ \sigma {Y}_{\mu\nu} +\alpha_5 {W}_{\mu\nu}}\right)=0,
\end{equation}
where 
\begin{eqnarray} \label{eqX-5d}
X_{\mu\nu}&=&  -\frac{1}{2} \left(\alpha {\cal L}_2 +\beta {\cal L}_3 \right) g_{\mu\nu} + \tilde X_{\mu\nu},\\
\tilde X_{\mu\nu}&=& {\cal K}_{\mu\nu} -[{\cal K}]g_{\mu\nu} -\alpha \left\{{{\cal K}_{\mu\nu}^2-[{\cal K}]{\cal K}_{\mu\nu} }\right\}  +\beta \left\{{{\cal K}_{\mu\nu}^3-[{\cal K}]{\cal K}_{\mu\nu}^2+\frac{{\cal L}_2}{2} {\cal K}_{\mu\nu} }\right\}, \\
\label{eqY-5d}
 Y_{\mu\nu} &=& -\frac{{\cal L}_4}{2} g_{\mu\nu} + \tilde Y_{\mu\nu},\\
\tilde Y_{\mu\nu} &=& \frac{{\cal L}_3}{2} {\cal K}_{\mu\nu}  -\frac{{\cal L}_2}{2}  {\cal K}^2_{\mu\nu} +[{\cal K}]{\cal K}^3_{\mu\nu} -{\cal K}^4_{\mu\nu},\\
\label{eqW-5d}
W_{\mu\nu} &=& -\frac{{\cal L}_5}{2}g_{\mu\nu} + \tilde W_{\mu\nu}, \\
\tilde W_{\mu\nu} &=& \frac{{\cal L}_4}{2}  {\cal K}_{\mu\nu} -\frac{{\cal L}_3}{2} {\cal K}^2_{\mu\nu}+\frac{{\cal L}_2}{2} {\cal K}^3_{\mu\nu} - [{\cal K}]{\cal K}^4_{\mu\nu} +{\cal K}^5_{\mu\nu}, 
\end{eqnarray}
with $\alpha = \alpha_3+1$, $\beta =\alpha_3+\alpha_4$, and $\sigma =\alpha_4+\alpha_5$ as additional parameters introduced for convenience. On the other hand, the corresponding constraint equations associated with the existence of reference metric turn out to be
\begin{equation}\label{5d-constraints}
t_{\mu\nu}\equiv \tilde X_{\mu\nu}+\sigma \tilde Y_{\mu\nu}+\alpha_5\tilde W_{\mu\nu} -\frac{1}{2} \left(\alpha_3 {\cal L}_2+\alpha_4 {\cal L}_3+ \alpha_5 {\cal L}_4 \right)g_{\mu\nu} = 0.
\end{equation}

According to a detailed investigation made in \cite{WFK}, the modified Einstein field equations (\ref{Einstein-5d}) can be reduced to
\begin{equation} \label{Einstein-5d-reduced}
\left({R_{\mu\nu}-\frac{1}{2}Rg_{\mu\nu}}\right) -\frac{m_g^2}{2}{\cal L}_M g_{\mu\nu}=0,
\end{equation}
with ${\cal L}_M \equiv {\cal L}_2+\alpha_3 {\cal L}_3 +\alpha_4 {\cal L}_4 +\alpha_5 {\cal L}_5$ as the total massive graviton Lagrangian. The equation (\ref{Einstein-5d-reduced}) implies an important result that the massive graviton terms act as an effective cosmological constant, $\Lambda_M \equiv - {m_g^2 {\cal L}_M}/2$, as a  consequence of the Bianchi identity, $\partial^\nu {\cal L}_M=0$. It has been shown in~\cite{TQD} that  this result turns out to be valid for a large class of physical and compatible reference metrics.  It appears that once the constraint equations (\ref{5d-constraints}), which are indeed algebraic rather than differential, are solved, the corresponding values of $\Lambda_M$ may be figured out~\cite{TQD}. 
%%%%%%%%%
\subsection{Five-dimensional massive bi-gravity}
Analogous to the five-dimensional massive gravity~\cite{TQD}, the corresponding action of five-dimensional massive bi-gravity is defined to be~\cite{TQD1}
\begin{eqnarray} 
&&S_{{5d,bi}} = M_g^2 \int {d^5 } x\sqrt {g} R(g)+ M_f^2 \int {d^5 } x\sqrt {f} R(f) \nonumber\\
&&+2m^2 M_{{eff}}^2 \int {d^5 } x\sqrt { g} \Bigl( {\cal U}_2 +\alpha_3 {\cal U}_3 +\alpha_4 {\cal U}_4 +\alpha_5 {\cal U}_5 \Bigr), 
\end{eqnarray}
with ${\cal U}_i ={\cal L}_i/2$ $(i=2-5)$ and therefore ${\cal U}_M\equiv{\cal U}_2 +\alpha_3 {\cal U}_3 +\alpha_4 {\cal U}_4 +\alpha_5 {\cal U}_5 ={\cal L}_M/2 $. As a result, the corresponding five-dimensional Einstein field equations turn out to be \cite{higher-bigravity,higher-bigravity-more,TQD1}
\begin{equation} \label{Einstein-5d-bi}
M_g^2 \left({R_{\mu\nu}-\frac{1}{2}Rg_{\mu\nu}}\right)+m^2 M_{{eff}}^2 {\cal H}_{\mu\nu}^{(5)} (g)=0,
\end{equation}
where 
\begin{eqnarray} \label{def-of-H}
{\cal H}_{\mu\nu}^{(5)} (g)\equiv{ { X}_{\mu\nu}+ \sigma {Y}_{\mu\nu} +\alpha_5 {W}_{\mu\nu}}.
\end{eqnarray}
Note that the tensors ${ X}_{\mu\nu}$, ${Y}_{\mu\nu}$, and ${W}_{\mu\nu}$ have been defined in Eqs. (\ref{eqX-5d}), (\ref{eqY-5d}), and (\ref{eqW-5d}). In fact, Eq. (\ref{Einstein-5d-bi}) turns out to be identical to Eq. (\ref{Einstein-5d}) if $M_f =0$, or equivalently $M_{eff} =M_g$. 

Now, for the reference metric, we can derive its field equations to be
\begin{equation} \label{5d-equation-f}
\sqrt {f} M_f^2\left(R_{\mu\nu}(f)-\frac{1}{2}f_{\mu\nu}R(f)\right)+ \sqrt{g} m^2 M_{{eff}}^2 s_{\mu\nu}^{(5)}(f) =0,
\end{equation}
where the tensor $s_{\mu\nu}^{(5)}(f) $ is given by
\begin{eqnarray} \label{5d-def-of-s}
s_{\mu\nu}^{(5)}(f) &\equiv& - {\hat{\cal K}}_{\mu\nu} +\Bigl\{ [{\cal K}]+ \alpha_3 {\cal U}_2 +\alpha_4 {\cal U}_3 +\alpha_5{\cal U}_4\Bigr\} f_{\mu\nu}  + \alpha \left\{{{\hat{\cal K}}_{\mu\nu}^2-[{\cal K}]{\hat{\cal K}}_{\mu\nu} }\right\} \nonumber\\
&& -\beta \left\{{{\hat{\cal K}}_{\mu\nu}^3-[{\cal K}]{\hat{\cal K}}_{\mu\nu}^2+{\cal U}_2 {\hat{\cal K}}_{\mu\nu} }\right\}  - \sigma \left\{{\cal U}_3 {\hat{\cal K}}_{\mu\nu}  - {\cal U}_2  {\hat{\cal K}}^2_{\mu\nu} +[{\cal K}]{\hat{\cal K}}^3_{\mu\nu} -{\hat{\cal K}}^4_{\mu\nu} \right\} \nonumber\\
&& -\alpha_5 \left\{ {\cal U}_4  {\hat{\cal K}}_{\mu\nu} - {\cal U}_3 {\hat{\cal K}}^2_{\mu\nu}+{\cal U}_2 {\hat{\cal K}}^3_{\mu\nu} - [{\cal K}]{\hat{\cal K}}^4_{\mu\nu}  +{\hat{\cal K}}^5_{\mu\nu} \right\}.
\end{eqnarray}
where the hatted tensors are defined as $\hat {\cal K}_{\mu\nu}= {\cal K}^{\sigma}_{\mu} f_{\sigma \nu}$, $\hat  {\cal K}_{\mu\nu}^2 = {\cal K}^{\rho}_{\mu}{\cal K}^{\sigma}_{\rho}f_{\sigma\nu}$, $\hat {\cal K}_{\mu\nu}^3= {\cal K}^{\rho}_{\mu}{\cal K}^{\delta}_{\rho}{\cal K}^{\sigma}_{\delta}  f_{\sigma\nu}$,  $\hat {\cal K}^4_{\mu\nu}={\cal K}^{\rho}_{\mu}{\cal K}^{\delta}_{\rho} {\cal K}^{\gamma}_{\delta} {\cal K}^{\sigma}_{\gamma}  f_{\sigma\nu}$, and ${\hat{\cal K}}^5_{\mu\nu} ={\cal K}^{\rho}_{\mu}{\cal K}^{\delta}_{\rho} {\cal K}^{\gamma}_{\delta} {\cal K}^{\sigma}_{\gamma}  {\cal K}^{\alpha}_{\sigma} f_{\alpha\nu}$ for convenience. It is clear that Eq. (\ref{5d-equation-f}) will no longer be algebraic but differential. Hence, it is not easy to have the constant-like behavior of graviton terms in the context of bi-gravity. Note, however, that there are two constraint equations due to the Bianchi identities for the physical and reference metrics in the five-dimensional bi-gravity~\cite{higher-bigravity,higher-bigravity-more,TQD1}
\begin{eqnarray} \label{constraint1}
D^\mu_g {\cal H}^{(5)}_{\mu\nu}(g)&=&0,\\
 \label{constraint2}
D^\mu_f \left[ \frac{\sqrt{g}}{\sqrt{f}}s_{\mu\nu}^{(5)}(f) \right]&=&0,
\end{eqnarray}
where we use notations $D^\mu_g$ and $D^\mu_f$ for the covariant derivatives in the physical and reference sectors, respectively. As a result, these two constraint equations (\ref{constraint1}) and (\ref{constraint2}) turn out be very useful in order to seek solutions because they lead the graviton terms to effective cosmological constants in the context of bi-gravity. It is worth noting that one of solutions making the graviton terms of four-dimensional bi-gravity constant is that the reference metrics are proportional to the physical ones~\cite{SFH,review-bigravity}. Hence, one might expect that this result will also hold in higher-dimensional scenarios of bi-gravity theory, especially the five-dimensional one.
%%%%%%%%%%%%%%%%%%%%%%%%%%%%%
\section{Simple cosmological solutions}\label{sec3}
Armed with the field and constraint equations of both five-dimensional massive gravity and bi-gravity, we are now able to seek their physical and cosmological solutions. Indeed, the Friedmann-Lemaitre-Robertson-Walker, Bianchi type I, and Schwarzschild-Tangherlini metrics have been shown to be solutions of five-dimensional massive gravity  and bi-gravity in \cite{TQD} and \cite{TQD1}, respectively. Now, we would like to remark some basic details of investigations carried out in these papers~\cite{TQD,TQD1}. 
%%%%%%%%%%%%%%%%%%%%%
\subsection{Friedmann-Lemaitre-Robertson-Walker metrics}
As a result,  we have considered the Friedmann-Lemaitre-Robertson-Walker spacetime~\cite{5d-FRW} for both physical and reference metrics of in both papers~\cite{TQD,TQD1}:
\begin{eqnarray} \label{frw-1}
&&ds_{{5d}}^2 (g) = -N_1^2(t)dt^2 + a_1^2(t) \left(d\vec{x}^2+du^2 \right), \\
 \label{frw-2}
&&ds_{{5d}}^2 (f) = -N_2^2(t)dt^2 + a_2^2(t) \left(d\vec{x}^2 +du^2 \right),
\end{eqnarray}
where  $u$ is the fifth dimension, while $a_1$ and $a_2$ are the scale factors of physical and reference metrics, respectively. Note that  the unitary gauge for the St\"uckelberg fields, i.e., $\phi^a =x^a$, is normally used in the context of massive gravity~\cite{RGT,review,TQD}, while the massive bi-gravity does not address the existence of St\"uckelberg fields~\cite{SFH,review-bigravity,TQD1}.

As a result,  we have been able to obtain the following ${\cal L}_M$ for the massive gravity as~\cite{TQD}
\begin{eqnarray}  \label{frw-9}
{\cal L}_M &=& 2 \Bigl[{\left({\gamma\alpha_4-\gamma_3}\right)\Sigma^3
+3\left({\gamma_2-\gamma\gamma_3}\right)\Sigma^2   +3 \left({\gamma\gamma_2-\gamma_1}\right) \Sigma -\gamma \gamma_1 +\left({3\gamma_1-3\gamma_2+\gamma_3}\right) } \Bigr]  \nonumber\\
&& + 2 \Bigl\{\alpha_4 \Sigma^3 -\alpha_5 \left(\gamma-1\right) \left(\Sigma-1\right)^3   -3\left[\gamma_3- \left(\gamma-1\right)\alpha_4 \right] \Sigma^2 +  3 \left[\gamma_2 -\left(\gamma -1\right)  \left(\gamma_3+\alpha_4 \right) \right] \Sigma   \nonumber\\
&&   + \left(\gamma-1\right) \left(3\gamma_3 +1 \right) -\gamma_1 \Bigr\} \left(\Sigma-1\right).
\end{eqnarray}
Note that additional parameters appearing in Eq. (\ref{frw-9}) are defined as 
\begin{eqnarray} \label{def-gamma}
&&\left[{\cal K} \right]^n = \left(5-\gamma-4\Sigma \right)^n;~\left[{\cal K}^n \right] = \left(1-\gamma \right)^n +4\left(1-\Sigma\right)^n;~\gamma = \frac{N_2}{N_1}; ~ \Sigma =\frac{a_2}{a_1};\nonumber\\
&&\gamma_1 = 3+3\alpha_3+\alpha_4; ~\gamma_2=1+2\alpha_3+\alpha_4;~ \gamma_3=\alpha_3+\alpha_4.
\end{eqnarray}
According to the method used in \cite{WFK}, we first solve the Euler-Lagrange equations of $N_2$ and $a_2$:
\begin{equation} \label{frw-Euler}
\frac{\partial {\cal L}_M}{\partial N_2}=0; ~ \frac{\partial {\cal L}_M}{\partial a_2}=0,
\end{equation}
which can be respectively reduced to the following equations:
\begin{equation} \label{frw-Euler-1}
\frac{\partial {\cal L}_M}{\partial \gamma}=0; ~ \frac{\partial {\cal L}_M}{\partial \Sigma}=0.
\end{equation}
Given the explicit definition of ${\cal L}_M$ in Eq. (\ref{frw-9}), the constraint equations (\ref{frw-Euler-1}) become
\begin{eqnarray}\label{frw-10}
 \alpha_5 \hat\Sigma^3+4\alpha_4 \hat\Sigma^2+6\alpha_3 \hat\Sigma+4 &=&0, \\
\label{frw-11}
 \left(\alpha_5 \hat\Sigma^3 +3\alpha_4 \hat\Sigma^2 +3\alpha_3 \hat\Sigma +1\right)\hat\gamma + \left(\alpha_4 \hat\Sigma^2+3\alpha_3 \hat\Sigma+3\right)\hat\Sigma &=&0,
\end{eqnarray}
where $\hat\Sigma \equiv 1-\Sigma$ and $\hat\gamma \equiv 1-\gamma$ as additional variables. Of course, these equations can also be obtained from  the  constraint equations $t_{\mu\nu}=0$ as shown in Eq. (\ref{5d-constraints}). It turns out that both Eqs. (\ref{frw-10}) and (\ref{frw-11}) are non-linear equations of $\hat\Sigma$. Hence, solving them will give us several values of $\hat\Sigma$ and therefore ${\cal L}_M$ (see \cite{TQD}  for more details). Once the value of ${\cal L}_M$ is figured out, the corresponding Einstein field equations (\ref{Einstein-5d-reduced}) can be solved to give
\begin{equation}
a_1(t) = \exp\left[\sqrt{\frac{\Lambda_M}{6}} t\right],
\end{equation} 
where $\Lambda_M \equiv - {m_g^2 {\cal L}_M}/2$. It is indeed the de Sitter expanding solution assuming that $\Lambda_M>0$.

Now, we turn to discuss the massive bi-gravity, whose physical metric equations are given by~\cite{TQD1}
\begin{eqnarray} \label{physical-eq-1}
6\tilde M_g^2 H_1^2 +\hat\Sigma\left(\sigma \hat\Sigma^3 + 4\beta \hat\Sigma^2 +6\alpha \hat\Sigma +4 \right)&=&0,\\
\label{physical-eq-2}
 3 \tilde M_g^2 \left(\dot H_1 +2 H_1^2 \right) +\hat\gamma \left(\sigma \hat\Sigma^3+3\beta \hat\Sigma^2+3\alpha \hat\Sigma +1\right) +\hat\Sigma \left(\beta \hat\Sigma^2+3\alpha \hat\Sigma +3\right)&=&0, 
\end{eqnarray}
along with its reference metric equations defined as
\begin{eqnarray} \label{reference-eq-1}
6\tilde M_f^2 \left(1- \hat\Sigma \right)^4 H_2^2  -\hat\Sigma \left(1-\hat\gamma\right)^2 \left(\alpha _5 \hat\Sigma ^3+4 \alpha _4 \hat\Sigma ^2+6 \alpha _3 \hat\Sigma +4\right) &=&0 ,\\
\label{reference-eq-2}
3 \tilde M_f^2 \left(1-\hat\Sigma\right)^3 \left(\dot H_2 +2 H_2^2 +\frac{\dot{\hat\gamma}}{1-\hat\gamma}H_2\right) &&\nonumber\\
  - \left(1-\hat\gamma \right)  \left[ \left(\alpha _5\hat\gamma+\alpha _4 \right)\hat\Sigma ^3 +3 \left(\alpha _4 \hat\gamma+\alpha _3 \right) \hat\Sigma ^2 +3  \left(\alpha _3 \hat\gamma +1\right)\hat\Sigma+\hat\gamma \right] &=& 0,
\end{eqnarray}
 where  $H_i \equiv \dot a_i/a_i$ as the Hubble constants for the physical and reference metrics. In addition, we have set additional variables as $\tilde M_g^2 \equiv M_g^2/(m^2 M_{{eff}}^2)$ and $\tilde M_f^2 \equiv M_f^2/(m^2 M_{{eff}}^2)$, while $\hat\gamma \equiv 1-N_2$ since $N_1$ has been set to be one. Besides these equations, we also have the other constraint equations coming from Eqs. (\ref{constraint1}) and (\ref{constraint2}):
\begin{eqnarray} \label{constraint3}
 \partial_0 \left[\hat\Sigma \left(\sigma \hat\Sigma^3 + 4\beta \hat\Sigma^2 +6\alpha \hat\Sigma +4 \right)\right] +4 H_1 \left(\hat\Sigma-\hat\gamma\right) \left(\sigma \hat\Sigma^3+3\beta \hat\Sigma^2+3\alpha \hat\Sigma +1\right) &=&0,\\
\label{constraint4}
  \left(4H_1 -4H_2 +\frac{3\dot{\hat\gamma}}{1-\hat\gamma} +\partial_0 \right)\left(1-\hat\gamma \right) \hat\Sigma \left(\alpha _5 \hat\Sigma ^3+4 \alpha _4 \hat\Sigma ^2+6 \alpha _3 \hat\Sigma +4\right)&& \nonumber\\
 + 4H_2 \left(\hat\Sigma-\hat\gamma\right) \left(\sigma \hat\Sigma^3+3\beta \hat\Sigma^2+3\alpha \hat\Sigma +1\right) &=&0.
\end{eqnarray}
As discussed in \cite{TQD1}, one of solutions to both equations (\ref{constraint3}) and (\ref{constraint4}) is that the reference metric is proportional to the physical one, i.e.,
\begin{equation}
f_{\mu\nu}= (1-{\hat C})^2  g_{\mu\nu}, 
\end{equation}
which is equivalent to $\hat\gamma=\hat\Sigma = \hat C$, where $\hat C$ is a constant obeying the following equation:
\begin{eqnarray} \label{eq-of-hat-C}
 &&\sigma \hat C^5 -2 \left(\sigma-2\beta \right) \hat C^4 + \left(\sigma-8\beta + 6\alpha + \alpha_5 \tilde M^2 \right)\hat C^3 + 4\left(\beta -3\alpha +\alpha_4 \tilde M^2+1\right)\hat C^2 \nonumber\\
 &&+2\left(3\alpha+3\alpha_3\tilde M^2-4\right) \hat C +4 \left( \tilde M^2+1 \right) =0,
\end{eqnarray}
with $\tilde M^2 \equiv \tilde M_g^2 / \tilde M_f^2$ as a dimensionless parameter.  It is noted that Eq. (\ref{eq-of-hat-C}) is a result of an equality, ${\hat\Lambda}^g_0=(1-\hat C)^2{\hat\Lambda}^f_0$, with ${\hat\Lambda}^g_0$ and ${\hat\Lambda}^f_0$ are effective cosmological constants of physical and reference sectors, whose values are defined as follows
\begin{equation}
{\hat\Lambda}^g_0 = -\frac{\hat C}{\tilde M_g^2}\left(\sigma \hat C^3 + 4\beta \hat C^2 +6\alpha \hat C +4 \right);~{\hat\Lambda}^f_0 = \frac{ \hat C}{ \tilde M_f^2 (1-\hat C)^4} \left(\alpha _5 \hat C ^3+4 \alpha _4 \hat C ^2+6 \alpha _3 \hat C +4\right).
\end{equation}
More interestingly, this solution leads to the constant-like feature of the graviton terms ${\cal L}_M$, which will make the field equations in both physical and reference sectors simpler to solve. It is noted that Eq. (\ref{eq-of-hat-C}) will reduce to
\begin{equation}
\alpha_5 \hat C^3 +4\alpha_4 \hat C^2 +6\alpha_3 \hat C +4=0,
\end{equation}
in the massive gravity limit, in which $s^{(5)}_{\mu\nu}=0$ since  the reference metric $f_{\mu\nu}$ is non-dynamical. Of course, this equation  is identical to Eq. (\ref{frw-10}) noting that $\hat C=\hat \Sigma$. As a result, solving the corresponding field equations of bi-gravity yields~\cite{TQD1} 
\begin{equation} \label{a1}
a_ 1 (t) = \exp \left[\sqrt{\frac{\hat\Lambda_0^g}{6}} t \right];~ a_2 (t) = (1- \hat C) \exp \left[\sqrt{\frac{\hat\Lambda_0^g}{6}} t \right].
\end{equation}

It is noted in the context of massive bi-gravity that ${\hat\Lambda}^g_0 M_g^2$ is different from $\Lambda_M$ derived from the massive graviton terms~\cite{TQD1} 
\begin{eqnarray} \label{lambdaM_expression}
\Lambda_M &=& -m^2 M_{{eff}}^2 \hat C \left[\left(\sigma \hat C^3 + 4\beta \hat C^2 +6\alpha \hat C +4 \right)+\left(\hat C-1\right)\left(\alpha _5 \hat C ^3+4 \alpha _4 \hat C ^2+6 \alpha _3 \hat C +4\right) \right] \nonumber\\
&=& \hat\Lambda_0^g \left[M_g^2 +M_f^2 (1-\hat C)^3\right].
\end{eqnarray}
%%%%%%%%%%%%%%%%%%%%%%%%%%%%%%%%%%
\subsection{Bianchi type I metrics}
This subsection is devoted to consider the five-dimensional Bianchi type I spacetime~\cite{5d-bianchi} for both physical and reference metrics of massive (bi-)gravity~\cite{TQD,TQD1}:
\begin{eqnarray}
\label{eq7}
 ds_{{5d}}^2 (g) &= & -N_1^2(t)dt^2+\exp\left[{2\alpha_1(t)-4\sigma_1(t)}\right]dx^2+\exp\left[{2\alpha_1(t)+2\sigma_1(t)}\right]\left({dy^2+dz^2}\right) \nonumber\\
&&+\exp\left[{2\beta_1(t)}\right] du^2 , \\
 \label{eq8}
 ds_{ {5d}}^2 (f)  &= &  -N_2^2(t)dt^2+\exp\left[{2\alpha_2(t)-4\sigma_2(t)}\right]dx^2+\exp\left[{2\alpha_2(t)+2\sigma_2(t)}\right]\left({dy^2+dz^2}\right)\nonumber\\
&&+\exp\left[{2\beta_2(t)}\right] du^2 .
\end{eqnarray}
As a result, the corresponding total graviton term ${\cal L}_M$ turns out to be~\cite{TQD,TQD1}
\begin{eqnarray} \label{explicit-LM}
{\cal L}_M  &= &2\Bigl[{\left({\gamma\alpha_4-\gamma_3}\right) AB^2 +\left({\gamma_2-\gamma\gamma_3}\right)B\left({2A+B}\right) }\nonumber\\ 
&& {+\left({\gamma\gamma_2-\gamma_1}\right)\left({A+2B}\right) -\gamma \gamma_1 +\left({3\gamma_1-3\gamma_2+\gamma_3}\right) }\Bigr]  \nonumber\\
&&+2 \Bigl\{ \alpha_4 AB^2 -\alpha_5 \left(\gamma-1\right) \left(A-1\right) \left(B-1\right)^2   - \left[\gamma_3- \left(\gamma-1\right)\alpha_4 \right] B \left(2A+B\right)  \nonumber\\
&&  +\left[\gamma_2 -\left(\gamma -1\right)  \left(\gamma_3+\alpha_4 \right) \right] \left(A+2B\right)   + \left(\gamma-1\right) \left(3\gamma_3 +1 \right) -\gamma_1 \Bigr\} \left(C-1\right),
\end{eqnarray}
where we have defined additional variables:
\begin{eqnarray}
A=\epsilon \eta^{-2}; ~ B=\epsilon \eta; ~C= \exp\left[\beta_2-\beta_1\right];~\epsilon = \exp\left[\alpha_2 -\alpha_1 \right]; ~ \eta =\exp\left[\sigma_2-\sigma_1\right].
\end{eqnarray}

First, we discuss the massive gravity.  Similar to the previous subsection for the FLRW metrics, we will first solve the Euler-Lagrange equations of scale factors of reference metric~\cite{TQD}:
 \begin{equation} \label{Euler}
\frac{\partial {\cal L}_M}{\partial N_2}=0; ~ \frac{\partial {\cal L}_M}{\partial \alpha_2}=0; ~\frac{\partial {\cal L}_M}{\partial \sigma_2}=0; ~\frac{\partial {\cal L}_M}{\partial \beta_2}=0;
\end{equation}
which are reduced to the following equations:
\begin{equation} \label{Euler-1}
\frac{\partial {\cal L}_M}{\partial \gamma}=0; ~ \frac{\partial {\cal L}_M}{\partial A}=0; ~\frac{\partial {\cal L}_M}{\partial B}=0; ~\frac{\partial {\cal L}_M}{\partial C}=0.
\end{equation}
As a result, these equations can be expanded to be non-linear equations of $A$, $B$, $C$, and $\gamma$~\cite{TQD}
\begin{eqnarray} \label{bianchi-constraint-1}
&&\alpha_4 AB^2-\gamma_3B\left({2A+B}\right) +\gamma_2\left({A+2B}\right)-\gamma_1 +\Bigl[  \alpha_4 B\left(2A+B\right) -\alpha_5 \left(A-1\right)\left(B-1\right)^2 \nonumber\\
&&  -\left(\gamma_3+\alpha_4\right) \left(A+2B\right) +3\gamma_3 +1 \Bigr] \left(C-1\right)=0, 
\end{eqnarray}
\begin{eqnarray}\label{bianchi-constraint-2}
&&\left({\gamma \alpha_4  -\gamma_3}\right)B^2+2\left({\gamma_2-\gamma \gamma_3}\right)B-\gamma_1+\gamma\gamma_2\nonumber\\
 &&+ \Bigl\{ \alpha_4 B^2  -\alpha_5 \left(\gamma-1\right) \left(B-1\right)^2- 2\left[\gamma_3 - \left(\gamma-1\right)\alpha_4\right]B    - \left(\gamma-1\right) \left(\gamma_3+\alpha_4\right) +\gamma_2 \Bigr\} \left(C-1\right)=0, \nonumber\\
\end{eqnarray}
\begin{eqnarray}\label{bianchi-constraint-3}
&&\left({\gamma \alpha_4  -\gamma_3}\right)AB +\left({\gamma_2-\gamma \gamma_3}\right)\left({A+B}\right)-\gamma_1+\gamma \gamma_2  +\Bigl\{ \alpha_4 AB-\alpha_5 \left(\gamma-1\right) \left(A-1\right)\left(B-1\right) \nonumber\\
&& - \left[\gamma_3 - \left(\gamma-1\right)\alpha_4\right] \left(A+B\right)    - \left(\gamma-1\right) \left(\gamma_3+\alpha_4\right) +\gamma_2 \Bigr\} \left(C-1\right)=0, 
\end{eqnarray}
\begin{eqnarray}\label{bianchi-constraint-4}
&&\alpha_4 AB^2-\gamma_3B\left({2A+B}\right) +\gamma_2\left({A+2B}\right)-\gamma_1 +\Bigl[  \alpha_4 B\left(2A+B\right) -\alpha_5 \left(A-1\right)\left(B-1\right)^2 \nonumber\\
&&  -\left(\gamma_3+\alpha_4\right) \left(A+2B\right) +3\gamma_3 +1 \Bigr] \left(\gamma-1\right)=0. 
\end{eqnarray}
As a result, solving these equations will yield several values of effective cosmological constant $\Lambda_M \equiv - {m_g^2 {\cal L}_M}/2$, see \cite{TQD} for a detailed investigation. Once the value of $\Lambda_M$ is figured out, we will have the corresponding component equations of Einstein field equations:
\begin{eqnarray}
 \label{bianchi-I-Einstein-2}
&&3\left(\dot\alpha_1^2 -\dot\sigma_1^2+\dot\alpha_1 \dot\beta_1  \right) = \Lambda_M; ~3\left(\ddot\alpha_1+2\dot\alpha_1^2 +\dot\sigma_1^2\right) =\Lambda_M;\\
 \label{bianchi-I-Einstein-4}
&&\ddot\sigma_1 + \dot\sigma_1 \left(3\dot\alpha_1 + \dot\beta_1 \right) =0; ~\ddot\beta_1 -2\dot\alpha_1^2+2\dot\sigma_1^2+\dot\beta_1^2+\dot\alpha_1 \dot\beta_1 =0.
\end{eqnarray}
It turns out that  solutions to these field equations can be found to be~\cite{TQD}
\begin{eqnarray}\label{bianchi-I-Einstein-12}
 \exp \left[3\alpha_1\right] &=& \exp\left[3\alpha_0\right] \left[\cosh \left(3\tilde H_1 t\right)+\frac{\dot\alpha_0}{\tilde H_1} \sinh \left(3\tilde H_1 t \right) \right], \nonumber\\
 \exp \left[\beta_1\right] &=& \exp\left[\beta_0\right] \left[\cosh \left(3\bar H_1 t\right)+\frac{\dot\beta_0}{3\bar H_1} \sinh \left(3\bar H_1 t \right) \right], \nonumber\\
\label{bianchi-I-Einstein-12}
\sigma_1 &=&\sigma_0 +\sqrt{\dot\alpha_0^2+\dot\alpha_0 \dot\beta_0 -H_1^2} \int \Biggl\{ \biggl[\cosh \left(3\tilde H_1 t\right)+\frac{\dot\alpha_0}{\tilde H_1} \sinh \left(3\tilde H_1 t \right) \biggr] \nonumber\\
&&  \times \biggl[\cosh \left(3\bar H_1 t\right)+\frac{\dot\beta_0}{3\bar H_1} \sinh \left(3\bar H_1 t \right) \biggr]\Biggr\}^{-1} dt, 
\end{eqnarray}
where $\alpha_0 \equiv \alpha_1(t=0)$, $\dot\alpha_0 \equiv \dot\alpha_1(t=0)$,  $\beta_0 = \beta_1(t=0)$, $\dot\beta_0 =\dot\beta_1(t=0)$, and $\sigma_0 =\sigma_1(t=0)$ are  initial values. In addition, some extra parameters existing in the above solutions can be understood as $\tilde H_1^2 =4 H_1^2/9(1-V_0)$,  $\bar H_1^2 = V_0 \tilde H_1^2$,  $H_1^2 \equiv \Lambda_M/3$, and $V_0$ is a constant  (see \cite{TQD} for more details).

Now, we would like to consider the five-dimensional massive bi-gravity to see whether it admits the above Bianchi type I metric as its solution. First, we examine the following constraint equations coming from the Bianchi identities (\ref{constraint1}) and (\ref{constraint2}):
\begin{eqnarray} \label{BI-constraint-1}
&&g^{00}\partial_0 {\cal H}^{(5)}_{00} -  g^{11} \left[\Gamma^{0}_{11}(g){\cal H}_{00}^{(5)}+\Gamma^{1}_{10}(g){\cal H}_{11}^{(5)}\right] - 2g^{22} \left[\Gamma^{0}_{22}(g){\cal H}_{00}^{(5)}+\Gamma^{2}_{20}(g){\cal H}_{22}^{(5)}\right] \nonumber\\
&&-g^{44} \left[\Gamma^{0}_{44}(g){\cal H}_{00}^{(5)}+\Gamma^{4}_{40}(g){\cal H}_{44}^{(5)}\right] =0
\end{eqnarray}
and
\begin{eqnarray}\label{BI-constraint-2}
&&f^{00} \left\{ \partial_0 \left[ \frac{\sqrt{g}}{\sqrt{f}}s_{00}^{(5)}\right] -2 \frac{\sqrt{g}}{\sqrt{f}}\Gamma^{0}_{00}(f) s_{00}^{(5)} \right\} - \frac{\sqrt{g}}{\sqrt{f}}f^{11} \left[ \Gamma^{0}_{11}(f)s_{00}^{(5)} +\Gamma^{1}_{10}(f) s_{11}^{(5)}\right] \nonumber\\
&&- 2\frac{\sqrt{g}}{\sqrt{f}}f^{22} \left[ \Gamma^{0}_{22}(f)s_{00}^{(5)} +\Gamma^{2}_{20}(f) s_{22}^{(5)}\right] - \frac{\sqrt{g}}{\sqrt{f}}f^{44} \left[ \Gamma^{0}_{44}(f)s_{00}^{(5)} +\Gamma^{4}_{40}(f) s_{44}^{(5)}\right] =0 .
\end{eqnarray}

As a result, we see that one of possible solutions to satisfy these constraint equations is that the reference metric $f_{\mu\nu}$ is proportional to the physical metric $g_{\mu\nu}$, similar to the FLRW case~\cite{TQD1}:
\begin{equation}
 f_{\mu\nu} =(1- {\tilde C})^2 g_{\mu\nu},
\end{equation}
 where  $ \tilde C=1-\gamma =1-A=1-B=1-C$ is a constant obeying the following equation:
\begin{eqnarray} \label{eq-of-tilde-C}
&&\sigma \tilde C^5 -2 \left(\sigma-2\beta \right) \tilde C^4 + \left(\sigma-8\beta + 6\alpha + \alpha_5 \tilde M^2 \right)\tilde C^3 + 4\left(\beta -3\alpha +\alpha_4 \tilde M^2+1\right)\tilde C^2 \nonumber\\ 
&&+2\left(3\alpha+3\alpha_3\tilde M^2-4\right) \tilde C +4 \left( \tilde M^2+1 \right) =0.
\end{eqnarray}
Furthermore, this solution also leads to the constant-like property of massive graviton term ${\cal L}_M$. Once the value of $\tilde C$ is solved, the corresponding value of effective cosmological constant  $\Lambda_M \equiv - {m_g^2 {\cal L}_M}/2$ can be determined. Thanks to this result, the following solutions can be defined to be~\cite{TQD1}
\begin{eqnarray}
 \exp[3\alpha_1] &=& \exp[3\alpha_{0}] \left[\cosh\left(3\tilde H_1 t\right)+\frac{\dot\alpha_{0}}{\tilde H_1}\sinh \left(3\tilde H_1 t\right)\right],\nonumber\\
 \exp[\beta_1] &=&\exp[\beta_{0}] \left[\cosh\left(3\bar H_1 t\right)+\frac{\dot\beta_{0}}{3\bar H_1}\sinh \left(3\bar H_1 t\right)\right],\nonumber\\
\label{bianchi-I-Einstein-22}
\sigma_1 &=&~\sigma_{0} +\sqrt{\dot\alpha_{0}^2+\dot\alpha_{0} \dot\beta_{0} -\frac{\tilde\Lambda_0^g}{3}} \int \Biggl\{ \biggl[\cosh \left(3\tilde H_1 t\right)+\frac{\dot\alpha_{0}}{\tilde H_1} \sinh \left(3\tilde H_1 t \right) \biggr] \nonumber\\
&& \times \biggl[\cosh \left(3\bar H_1 t\right)+\frac{\dot\beta_{0}}{3\bar H_1} \sinh \left(3\bar H_1 t \right) \biggr]\Biggr\}^{-1} dt,
\end{eqnarray}
where $\tilde H_1^2 = 4\tilde \Lambda_0^g/27(1-V_0^g)$, $\bar H_1^2= V_0^g \tilde H_1^2$, $V_0^g$ is a constant, and $\tilde \Lambda_0^g$ is an effective cosmological constant of physical sector, whose definition is given by
\begin{equation}
\tilde \Lambda_0^g = - \frac{\tilde C}{\tilde M_g^2} \left( \sigma \tilde C^3 +4\beta \tilde C^2 +6\alpha \tilde C +4 \right).
\end{equation}
It is noted that $\Lambda_M$ existing in Eq. (\ref{bianchi-I-Einstein-12}) is different from $\tilde \Lambda_0^g$ appearing in Eq. (\ref{bianchi-I-Einstein-22}) due to the dynamical property of reference metric of bi-gravity. It is worth noting that the discussions on the stability of Bianchi type I solutions of both massive gravity and bi-gravity have been written in \cite{TQD,TQD1}. 
%%%%%%%%%%%%%%%%%%%%%%%%%%%%%%%%%%%%%
\subsection{Schwarzschild-Tangherlini metrics}
In this third subsection, we will summarize main results investigated in \cite{TQD,TQD1} for the Schwarzschild-Tangherlini metrics of the following form~\cite{5d-sch}:
\begin{eqnarray} \label{five-dim-metric}
  ds_{{5d}}^2 (g) &= & -N_1^2\left(r\right) dt^2 +\frac{dr^2}{F_1^2\left(r\right)}+\frac{r^2 d\Omega_3^2}{H_1^2\left(r\right)},\\
 ds_{{5d}}^2 (f)&= & -N_2^2\left(r\right) dt^2 +\frac{dr^2}{F_2^2\left(r\right)}+\frac{r^2 d\Omega_3^2}{H_2^2\left(r\right)}, 
\end{eqnarray}
with $d\Omega_3^2=d\theta^2 +\sin^2 \theta d\varphi^2 + \sin^2 \theta \sin^2 \varphi d \psi^2$. As a result, the corresponding ${\cal L}_M$ becomes
\begin{eqnarray} \label{Lagra-reduced}
{\cal L}_M &=&2\left\{ \left[\alpha_5 \left({\cal K}^2{ }_2\right)^3+ 3\alpha_4 \left({\cal K}^2{ }_2\right)^2 +3 \alpha_3 {\cal K}^2{ }_2 +1\right] {\cal K}^0{ }_0 {\cal K}^1{ }_1  \right.\nonumber\\ 
&& \left.+{\cal K}^2{ }_2 \left[ \alpha_4 \left({\cal K}^2{ }_2\right)^2+3\alpha_3 {\cal K}^2{ }_2 +3\right] \left({\cal K}^0{ }_0 + {\cal K}^1{ }_1\right) + \left({\cal K}^2{ }_2\right)^2 \left(\alpha_3{\cal K}^2{ }_2+3 \right)\right\},
\end{eqnarray}
where 
\begin{equation}
{\cal K}^0{ }_0(r) =1- \frac{N_2}{N_1};~ {\cal K}^1{ }_1(r) = 1- \frac{F_1}{F_2};~{\cal K}^2{ }_2(r)= {\cal K}^3{ }_3(r) = {\cal K}^4{ }_4(r)=1-\frac{H_1}{H_2}.
\end{equation}
Similar to the previous subsections, we first consider the massive gravity by deriving its constraint equations ${\partial {\cal L}_M}/{\partial {\cal K}^0{ }_0} = {\partial {\cal L}_M}/{\partial {\cal K}^1{ }_1} = {\partial {\cal L}_M}/{\partial {\cal K}^2{ }_2} =0$ to be
\begin{eqnarray} %\label{constraint1}
\left(\alpha_5 {\cal K}^1{ }_1 +\alpha_4 \right)\left({\cal K}^2{ }_2\right)^3 +3 \left(\alpha_4 {\cal K}^1{ }_1 + \alpha_3 \right)\left({\cal K}^2{ }_2\right)^2 +3 \left(\alpha_3 {\cal K}^1{ }_1 +1 \right){\cal K}^2{ }_2+{\cal K}^1{ }_1 &=&0, \\
%\label{constraint4}
\left(\alpha_5 {\cal K}^0{ }_0 +\alpha_4 \right)\left({\cal K}^2{ }_2\right)^3 +3 \left(\alpha_4 {\cal K}^0{ }_0 + \alpha_3 \right)\left({\cal K}^2{ }_2\right)^2 +3 \left(\alpha_3 {\cal K}^0{ }_0 +1 \right){\cal K}^2{ }_2+{\cal K}^0{ }_0 &=&0, \\
\label{constraint5}
 \left[\alpha_5 \left({\cal K}^2{ }_2\right)^2 +2\alpha_4 {\cal K}^2{ }_2 +\alpha_3 \right] {\cal K}^0{ }_0 {\cal K}^1{ }_1+ \left[\alpha_4 \left({\cal K}^2{ }_2\right)^2 +2 \alpha_3 {\cal K}^2{ }_2 +1 \right] \left({\cal K}^0{ }_0+{\cal K}^1{ }_1\right)&& \nonumber\\
+ \left(\alpha_3 {\cal K}^2{ }_2+2\right){\cal K}^2{ }_2 &=&0.
\end{eqnarray}
As a result, solving these non-linear equations of ${\cal K}^0{ }_0$, ${\cal K}^1{ }_1$, and ${\cal K}^2{ }_2$ will imply several values of ${\cal L}_M$ (see \cite{TQD} for more details). Thanks to the constant-like behavior of ${\cal L}_M$, the physical Einstein field equations will turn out to be of the usual form with an effective cosmological constant $\Lambda_M \equiv - {m_g^2 {\cal L}_M}/2$. Indeed, the five-dimensional physical metric (\ref{five-dim-metric}) can be solved to be~\cite{5d-sch,TQD}
\begin{eqnarray} \label{five-dim-metric-1}
&&ds^2 = -f(r)dt^2 +\frac{dr^2}{f(r)}+r^2d\Omega_3^2, \nonumber\\
&& N_1^2(t,r)=F_1^2(t,r)=f(r)=1-\frac{\mu}{r^2}-\frac{\Lambda_M}{6}r^2,~H_1^2(t,r)=1, 
\end{eqnarray}
where $\mu \equiv {8 G_5 M}/({3\pi})$ the mass parameter, $M$ the mass of source, and $G_5$  the 5-dimensional Newton constant. It appears that if $\Lambda_M$ is positive  or negative definite then the metric (\ref{five-dim-metric-1}) will be called the  Schwarzschild-Tangherlini-de Sitter or Schwarzschild-Tangherlini-anti-de Sitter black hole, respectively. Otherwise,  we will have the pure Schwarzschild-Tangherlini black hole for $\Lambda_M=0$. 

For the five-dimensional massive bi-gravity, its Bianchi constraint equations are given by
\begin{eqnarray}\label{constraint-physical-Sch-1}
&&g^{11}\left[\partial_r {\cal H}^{(5)}_{11}-2\Gamma^1_{11}(g){\cal H}^{(5)}_{11}\right] - g^{00}\left[\Gamma^1_{00}(g) {\cal H}^{(5)}_{11} +\Gamma^0_{01}(g){\cal H}^{(5)}_{00} \right]-g^{22} \left[\Gamma^1_{22}(g){\cal H}^{(5)}_{11} +\Gamma^2_{21}(g){\cal H}^{(5)}_{22} \right]\nonumber\\
&&-g^{33} \left[\Gamma^1_{33}(g){\cal H}^{(5)}_{11} +\Gamma^3_{31}(g){\cal H}^{(5)}_{33} \right] -g^{44} \left[\Gamma^1_{44}(g){\cal H}^{(5)}_{11} + \Gamma^4_{41}(g){\cal H}^{(5)}_{44} \right] =0, \\
&&g^{33} \left[\Gamma^2_{33}(g){\cal H}^{(5)}_{22} + \Gamma^3_{32}(g){\cal H}^{(5)}_{33}\right]  + g^{44} \left[\Gamma^2_{44}(g){\cal H}^{(5)}_{22} + \Gamma^4_{42}(g){\cal H}^{(5)}_{44}\right]=0, \\
&&g^{44} \left[\Gamma^3_{44}(g){\cal H}^{(5)}_{33} + \Gamma^4_{43}(g){\cal H}^{(5)}_{44}\right] =0,
\end{eqnarray}
along with
\begin{eqnarray}\label{constraint-reference-Sch-1}
&&f^{11}\left\{ \partial_r \left[ \frac{\sqrt{g}}{\sqrt{f}}s^{(5)}_{11}\right]-2\frac{\sqrt{g}}{\sqrt{f}}\Gamma^1_{11}(f)s^{(5)}_{11}\right\} - \frac{\sqrt{g}}{\sqrt{f}} f^{00}\left[\Gamma^1_{00}(f) s^{(5)}_{11} +\Gamma^0_{01}(f)s^{(5)}_{00} \right] \nonumber\\
&&- \frac{\sqrt{g}}{\sqrt{f}}f^{22} \left[\Gamma^1_{22}(f)s^{(5)}_{11} +\Gamma^2_{21}(f)s^{(5)}_{22} \right]-\frac{\sqrt{g}}{\sqrt{f}}f^{33} \left[\Gamma^1_{33}(f)s^{(5)}_{11} +\Gamma^3_{31}(f)s^{(5)}_{33} \right]  \nonumber\\
&& -\frac{\sqrt{g}}{\sqrt{f}}f^{44} \left[\Gamma^1_{44}(f)s^{(5)}_{11} + \Gamma^4_{41}(f)s^{(5)}_{44} \right] =0,\\
&& f^{33} \left[\Gamma^2_{33}(f)s^{(5)}_{22} + \Gamma^3_{32}(f)s^{(5)}_{33}\right] + f^{44} \left[\Gamma^2_{44}(f)s^{(5)}_{22} + \Gamma^4_{42}(f)s^{(5)}_{44}\right] =0, \\
&& f^{44} \left[\Gamma^3_{44}(f)s^{(5)}_{33} + \Gamma^4_{43}(f)s^{(5)}_{44}\right] =0.
\end{eqnarray}

As a result, we observe that one of simple solutions to the  is that the physical and reference metrics are proportional to each other \cite{TQD1}, i.e.,
\begin{equation}
f_{\mu\nu} =(1 -\bar C)^2 g_{\mu\nu},
\end{equation}
where $\bar C$ is a constant obeying the following equation:
\begin{eqnarray} \label{eq-of-bar-C}
&&\sigma \bar C^5 -2 \left(\sigma-2\beta \right) \bar C^4 + \left(\sigma-8\beta + 6\alpha + \alpha_5 \tilde M^2 \right)\bar C^3 + 4\left(\beta -3\alpha +\alpha_4 \tilde M^2+1\right)\bar C^2 \nonumber\\ 
&&+2\left(3\alpha+3\alpha_3\tilde M^2-4\right) \bar C +4 \left( \tilde M^2+1 \right) =0.
\end{eqnarray}
Similar to the massive gravity, solving the corresponding Einstein field equations of physical metric gives \cite{5d-sch,TQD1}
\begin{eqnarray}% \label{five-dim-metric-1}
&&g^{{5d}}_{\mu\nu}dx^{\mu}dx^{\nu} = -f(r)dt^2 +\frac{dr^2}{f(r)}+r^2d\Omega_3^2, ~ N_1^2(r)=F_1^2(r)=f(r)=1-\frac{\mu}{r^2}-\frac{\bar\Lambda_0^g}{6}r^2,\nonumber\\
&&H_1^2(r)=1,~{\bar \Lambda}_0^g = -\frac{\bar C}{{\tilde M}^2_g} \left(\sigma \bar C^3+4\beta \bar C^2+6\alpha \bar C +4 \right).
\end{eqnarray}
%Once again, we should note that $\Lambda_M \neq \bar\Lambda_0^g$ in the context of massive bi-gravity due the dynamical property of reference metric.
%%%%%%%%%%%%%%%%%%%%%%%%%%%%%
\section{Conclusions}\label{sec4}
We have mainly summarized our results on both five-dimensional massive gravity and bi-gravity, which have been  published in two recent papers in \cite{TQD,TQD1}.  It turns out that both five-dimensional massive and bi-gravity are indeed physically non-trivial theories with a should-not-be-ignored ${\cal L}_5$ term, which disappears in all four-dimensional spacetimes but does exist in any five- (or higher) dimensional one. As a result, we have shown the useful method based on the Cayley-Hamilton theorem to construct arbitrary dimensional graviton terms ${\cal L}_i$ and therefore the five-dimensional graviton term ${\cal L}_5$. The following field equations of both five-dimensional physical and  reference metrics have been derived consistently in order to discuss cosmological aspects. Note that the field equations of reference metric have been regarded as the constrained equations, which should be solved first to determine the corresponding values of massive graviton terms. It turns out that by counting the contribution of ${\cal L}_5$, the value(s) of effective cosmological constants coming from massive graviton terms ${\cal L}_M$ will  be changed, and therefore the corresponding cosmological solutions will be different from that defined in scenarios without ${\cal L}_5$~\cite{WFK}. To end this conclusion section, we would like to note again that ${\cal L}_5$ should not be ignored in all higher-than-four dimensional massive (bi-)gravity. 
%%%%%%%%%%%%%%%%%
\ack This research is supported in part by VNU University of Science, Vietnam National University, Hanoi.

\end{document}